\documentclass[12pt]{iopart}
\usepackage{txfonts}
\usepackage{graphicx}
\usepackage{iopams,setstack}
\usepackage{cite}
\begin{document}
\title[]{The ``weak'' interdependence of infrastructure systems produces mixed percolation transitions in multilayer networks}

\author{Run-Ran Liu$^{1,2}$, Daniel A. Eisenberg$^3$, Thomas P. Seager$^3$, and Ying-Cheng Lai$^{2,4}$}
\address{$^1$ Alibaba Research Center for Complexity Sciences, Hangzhou Normal University, Hangzhou, Zhejiang 311121, China}
\address{$^2$ School of Electrical, Computer, and Energy Engineering, Arizona State University, Tempe, AZ 85287, USA}
\address{$^3$ School of Sustainable Engineering and Built Environment, Arizona State University, Tempe, AZ 85287, USA}
\address{$^4$ Department of Physics, Arizona State University, Tempe, AZ 85287, USA}

\date{\today}

\begin{abstract}
The robustness of individual networks depends upon their interdependent counterparts, as initiating failures can cascade across diverse systems. Understanding the ways in which interdependent links across multilayer networks influence systemic failures is a crucial first step to develop broadly influential resilience recommendations for real-world infrastructure. However, previous studies model cascading failures via a node-to-node percolation process that assumes ``strong'' interdependence across layers--once a node in any layer fails, its neighbors in other layers fail immediately and completely with all links removed. This assumption is not true of real interdependent infrastructures that have design standards and emergency procedures to buffer against cascades. In this work, we propose an interdependent, multilayer network model and percolation process that matches infrastructures better than previous models by allowing some nodes to survive when their interdependent neighbors fail. We consider a node-to-link failure propagation mechanism and establish ``weak'' interdependence across layers via a tolerance parameter $\alpha$ which quantifies the likelihood that a node survives when one of its interdependent neighbors fails. We measure the robustness of any individual layer by the final size of its giant component. Analytical and numerical results show that weak interdependence produces a striking phenomenon: layers at different positions within the multilayer system experience distinct percolation transitions. Especially, layers with high super degree values percolate in an abrupt manner, while those with low super degree values exhibit both continuous and abrupt transitions. This novel phenomenon we call \emph{mixed percolation transitions} has significant implications for network robustness. Previous results that do not consider cascade tolerance and layer super degree may be under- or over-estimating the vulnerability of real systems. Moreover, since $\alpha$ represents a generic measure of various risk management strategies used to buffer infrastructure assets from cascades, our model reveals how nodal protection activities influence failure dynamics in interdependent, multilayer systems.
\end{abstract}

\maketitle

\section{Introduction} \label{sec:intro}

The robustness of a complex networked system to survive random component failures and/or intentional attacks is a significant problem with broad implications. Robustness, i.e., the likelihood that a network remains functional after losing nodes or links, has been a topic of interest since the beginning of modern network science~\cite{AJB:2000,CEAH:2000,CEAH:2001,CNSW:2000,ML:2002,Watts:2002,HK:2002,MGP:2002,Holme:2002,MPSVV:2003,GGKK:2003,VSS:2004,CLM:2004,ZPL:2004,PSS:2005,ZPLY:2005,HLC:2008,YWLC:2009,YCL:2005}. In recent years, the robustness of interdependent, multilayered networked systems (sometimes referred to as networks-of-networks) has become a subject of focused study~\cite{PBH:2010,BPPSH:2010,PBH:2011,KABGP:2014,Du:2016}. Examples of real-world interdependent, multilayer networks are ample, with a common and important case being urban infrastructure systems~\cite{Ouyang:2014} such as transportation, communication, electric power, and water supply networks. The sharing of services within and between these infrastructure systems suggests that the loss of a single service like mobility can impact the provision of others like electric power and clean water -- when a node or link in one infrastructure network fails, because of dependencies across infrastructures, its neighbors in other networks can also fail~\cite{Rinaldi:2001}. Since failures can propagate from one network to others, understanding the ways in which dependencies within and across these multilayer networks influences systemic robustness is a crucial first step to develop broadly influential resilience recommendations.
The goal of this paper is to develop a model for the failure dynamics which is more realistic to interdependent infrastructure contexts than those in previous studies. To make the terminologies unambiguous, we use the multilayer network lexicon developed by Kivela \emph{et al}.~\cite{KABGP:2014} to describe our approach.

The phenomenon of failure propagation or spreading in a multilayer network depends on the dynamical or physical processes specific to the type of node or link failures. It is difficult to develop a general, dynamics based framework to address the robustness and resilience of systems like urban infrastructure where nodes, links, and layers may represent characteristically different objects and relationships. A viable approach is then to focus on the structural properties of the multilayer network through percolation. Indeed, percolation models~\cite{Broadbent:1957,Kirkpatrick:1973,Kesten:1982,Stauffer:1992}
have been used as a theoretical tool to quantify the robustness of networks subject to random failures~\cite{BPPSH:2010,Hackett:2016} or malicious
attacks~\cite{Huang:2011,Shao:2015}, where the robustness of the network as a function of the number of damaged elements can be characterized as
a phase transition. For interdependent, multilayer networks, a collapse can occur abruptly, such that the size of the mutually connected component changes to zero discontinuously upon removal of a critical fraction of nodes~\cite{BPPSH:2010,PBH:2010}. This result was somewhat surprising because it is characteristically different from the continuous percolation transition that is typical of single-layer networks. The abrupt and discontinuous nature of the percolation transition is a unique manifestation of interdependent nodes linked across layers, because damage to one network can spread to others. As the fraction of the interdependent nodes among the networks is reduced, a change from a discontinuous to continuous percolation transition
can occur~\cite{Parshani:2010,Gao:2011,Gao:2012,Bianconi:2014,Havlin:2015}. The way in which interdependent links are defined and generated also influences multilayer robustness properties, including coupling models based on inter-similarity~\cite{Parshani:2011,Hu:2013}, link overlap~\cite{Cellai:2013},
degree correlations~\cite{Zhou:2014,Valdez:2013,Min:2014}, clustering~\cite{Shao:2014,Huang:2013}, degree distribution~\cite{Emmerich:2014,Yuan:2015}, inner-dependency~\cite{Liu:2016A}, intersection among the network layers~\cite{Radicchi:2015}, and spatially embedded networks~\cite{Li:2012,Bashan:2013,Shekhtman:2014,Danziger:2014}. Depending on the types of interactions among nodes across network layers, some interdependent networks can also produce different types of percolation transitions, such as $k$-core~\cite{Azimi:2014}, weak~\cite{Baxter:2014}, and redundant~\cite{Radicchi:2017} percolations.

A tacit assumption in all of these percolation models is that interdependence within and across layers is ``strong'', which is not true of many real-world systems. A percolation model assumes strong interdependence when the loss of a node in one layer will always cause the removal of its neighbors and neighboring links across all interdependent layers, i.e., an inter-layer cascade probability of one. This assumption ignores important aspects of interdependent systems that provide buffers against cascades and protect layers from each other. For example, travel within and between cities is enabled by several interdependent modes of transportation like personal cars, buses, trains, airplanes, and ferries. When one mode becomes unavailable, e.g., the airport is shut down, this loss may increase travel demand for other modes within and between cities as people traveling on planes switch to using other means to reach their final destination. In this situation, the cascading failure processes with ``strong'' interdependence found in all previous models assumes that local travel via cars, buses, trains, and ferries and city-to-city travel via airports will be affected equally (removing all nodes and links). In reality, the likelihood that a failure will propagate from air travel to other modes and cities is unknown as each interdependent mode may have sufficient additional capacity to support increased demand. Thus, the failure of a node in one layer can disable a number of links in other coupled layers, but not necessarily cause the loss of all neighboring nodes and links. Assuming a node-to-node failure mechanism without any tolerance to cascades effectively ignores redundant infrastructures and adaptive practices used by transportation infrastructure providers to handle these situations. This notion of cascade tolerance influences the resilience of many real-world infrastructure systems, as electric power, water, transportation, and communication systems use a number of backup infrastructures and emergency management plans to survive losses of interdependent services. Instead of a node-to-node failure mechanism, cascade tolerance is captured by a node-to-link failure mechanism corresponding to bond percolation dynamics in percolation theory where nodes posses a probabilistic tolerance or susceptibility to nodal failures originating in a different layer. Despite the practical significance of bond percolation to understand failure dynamics in interdependent infrastructure systems~\cite{Reis:2014,Hu:2017PNAS}, the effects of node-to-link failure propagation on the robustness of a multilayer network has not been studied.

In this paper, we apply bond percolation dynamics to multilayer networks to uncover and understand the effects of nodal susceptibility on the robustness of interdependent systems. We assume the existence of buffers between interdependent network layers, such that the failure of a node will not lead to the failures of its interdependent neighbors in another layer with certainty, but with a probability that depends on the link connectivity. We denote the assumption of possible node and link survival in percolation models as assuming ``weak" interdependence. We capture weak interdependence by introducing a tolerance parameter $\alpha$ to quantify the heterogeneity of the probabilities. According to percolation theory, the robustness of a network in the whole multilayer networked system against failures can be characterized by the size of the final giant component after the process of failure propagation ends. Treating $\alpha$ as an external control parameter, we find that its tuning can lead to a series of phase transitions. In particular, multiple percolation transitions occur for relatively large values of $\alpha$, the networks in the system percolate continuously one after another according to their super degree values. However, for moderate values of $\alpha$, the phenomenon of mixed percolation transitions occurs, where networks with large super degrees in the interdependent system exhibit abrupt (first-order) percolation transitions while those with small super degree values display double transitions: one continuous (second-order)
transition followed by a discontinuous (first-order) transition. For relatively small values of $\alpha$, the percolation transition points merge together, leading to simultaneous and abrupt transition for all networks in the system. We obtain these results analytically and numerically, and they provide general insights into the robustness of multilayer networks and interdependent infrastructure systems.

\section{Model} \label{sec:model}

\begin{figure}
\centering
\includegraphics[width=\linewidth]{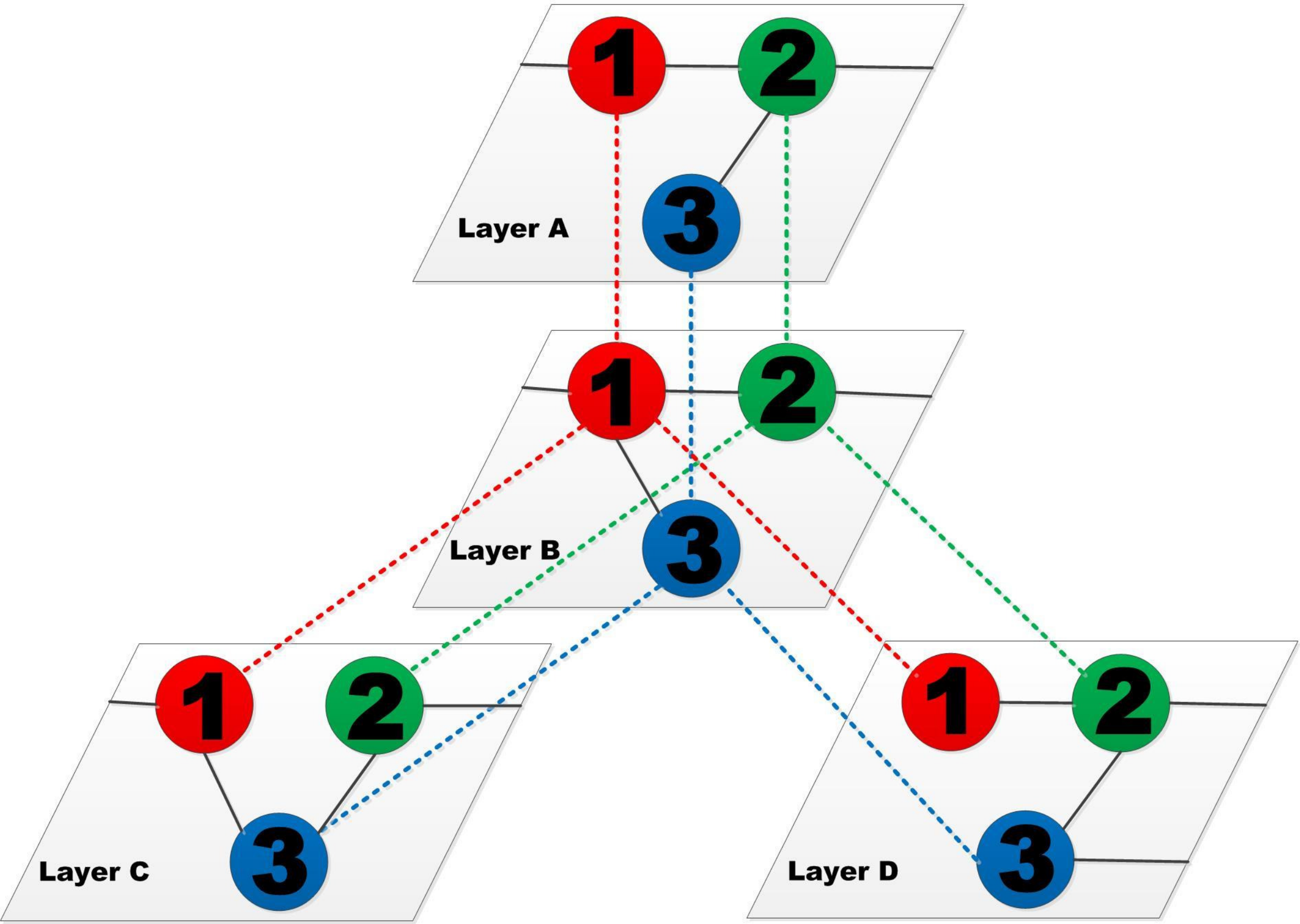}
\caption{ {\bf Schematic illustration of the structure of an interdependent networked system, or a network of networks}. The whole system comprises four layers of networks, where the color dotted lines denote the interdependent links among different networks and the black solid lines denote the connections inside a network.}
\label{fig:structure}
\end{figure}

We consider a percolation process on a system of $M$ layers of networks $A, B, C,\cdot\cdot\cdot$, each having the same number $N$ of nodes. The
networks form an interdependent, multilayer system with a possible structure shown in Fig.~\ref{fig:structure}. A node in the system can be denoted by a
pair of labels $(x, X)$, with $x (x=1,2,\cdot\cdot\cdot N)$ and $X (X=A,B,C\cdot\cdot\cdot)$ being the nodal and layer indices, respectively.
Nodes within the same layer $X$ are connected with the degree distribution $p_{k}^{X}$. Nodes in different layers sharing a common nodal index are
the replica nodes connected by dependency links if the underlying networks to which they belong are interdependent upon each other. That is, for every nodal label $x$, the set of $M$ replica nodes corresponding to pairs of labels $(x, X)$ can also be regarded as possessing a network structure through the dependency links. The number of interdependent networks, $q^{X}$, for network $X$ is the ``super degree,'' where every node in $X$ has $q^{X}$ interdependent replica nodes.

A randomly connected network layer $X$ with degree distribution $p_{k}^{X}$ is essentially a set of connected components~\cite{Molloy:1995}, where the nodes in the giant component are regarded as functional or viable and the remaining nodes are treated as failed or inviable. Our node-to-link failure propagation process can be described, as follows. Consider a pair of interdependent networks: $A$ and $B$. If a node $(x, A)$ in $A$ fails, each link of its replica $(x, B)$ in network $B$ will be disabled with probability $1-\alpha$. Similarly, if a node $(y, B)$ in network $B$ fails, the links of its replica $(y, A)$ in network $A$ will be cut off with the same probability. A initial failure due to isolation from the giant component in a certain network can then spread across the whole system through an iterative process. In each iteration, disconnecting certain nodes from the giant component of network $A$ will cause some nodes to be isolated from the giant component of network $B$, which in turn will induce more nodal failures in $A$. This recursive or cascading process occurs in all pairs of interdependent networks synchronously at each iteration. When no failures are possible, the whole system reaches a stable steady state.

From the perspective of a single node, the parameter $\alpha$ quantifies the tolerance to its failed interdependent replicas, which controls the impacts that a node will endure if one of its replicas fails and represents the probability that interdependent systems will be unsuccessful at preventing inter-layer cascades. From the standpoint of the whole networked system, $\alpha$ determines effectively the strength of interdependence among the networks. For $\alpha \rightarrow 1$, the interdependence between nodes is the weakest so that, practically, failures cannot spread from one network to another. The opposite limit $\alpha \rightarrow 0$ signifies the case where interdependence is the strongest. In this case, our model reduces to previous percolation models with the node-to-node failure propagation mechanisms~\cite{Gao:2011}. In our study, we use the size of the giant components $S^{X}$ for the final network layers $X$ $(X=A,B,\cdot\cdot\cdot)$ to measure the robustness of the system, as in previous works~\cite{BPPSH:2010}.

We consider our percolation model to be a simplified representation of urban infrastructure systems with buffers against interdependent infrastructure failures. A representative system following the general form of our model is multi-modal transportation linking multiple cities. To convert Fig.~\ref{fig:structure} into a transportation model, we would represent each colored node as a separate city and each layer as a different mode of transportation. Within cities, people travel via urban transportation (e.g., cars and buses) to different kinds of hub locations (e.g., coach stations, airports, and railway stations) to facilitate mode switching and interactions. One transportation layer is then linked via one kind of hubs in cities, as connecting highways, planes or railways link urban regions separated by large geographic distances. Although the current model is insufficient to consider specific travel dynamics within each city, our model provides a general form to approach the question of how dynamic buffers may lead to cascades within and across modes of transportation. In particular, the tuning of $\alpha$ from weak to strong interdependence captures situations where the hubs in different modes of travel have either an excess or a lack of capacity to handle the congestion caused by additional passengers, respectively. This model is general, such that cascade tolerance can be due to physical constraints (e.g., number of highway lanes), temporal constraints (e.g., early morning vs. rush hour traffic), and/or dynamic actions (e.g., lane shifting and emergency procedures). Thus, although simplified, robustness analysis of the multilayer system is informative to the ways in which transportation across urban regions may cascade and disconnect interdependent traffics. Moreover, the general form of our model can also capture buffers across other systems by treating layers as different infrastructures that link to transportation hubs within and across cities (e.g., electric power transmission substations).

\section{Theory} \label{sec:theory}

\subsection{General formalism}

We solve our model analytically in terms of the final state after a cascading process using the standard method of generating functions~\cite{Newman:2002,Son:2012,Son:2011}. Let $R^{X}$ be the probability that a randomly chosen link in network layer $X$ belongs
to the giant component, for $X\in\{A, B, C,\cdot\cdot\cdot\}$. The function $G^{X}_{0}(x)=\sum_kp_{k}^{X}x^k$ is the generating function that gives rise
to the degree distribution for random nodes in layer $X$, and $G^{X}_{1}(x)=\sum_{k}p_{k}^{X}kx^{k-1}/\langle k\rangle^X$ is the generating function for the underlying branching processes of layer $X$, which generates the distribution for the number of outgoing links of randomly chosen links in layer $X$, where $\langle k\rangle^X$ is the average degree of network $X$.

\begin{figure}
\centering
\includegraphics[width=\linewidth]{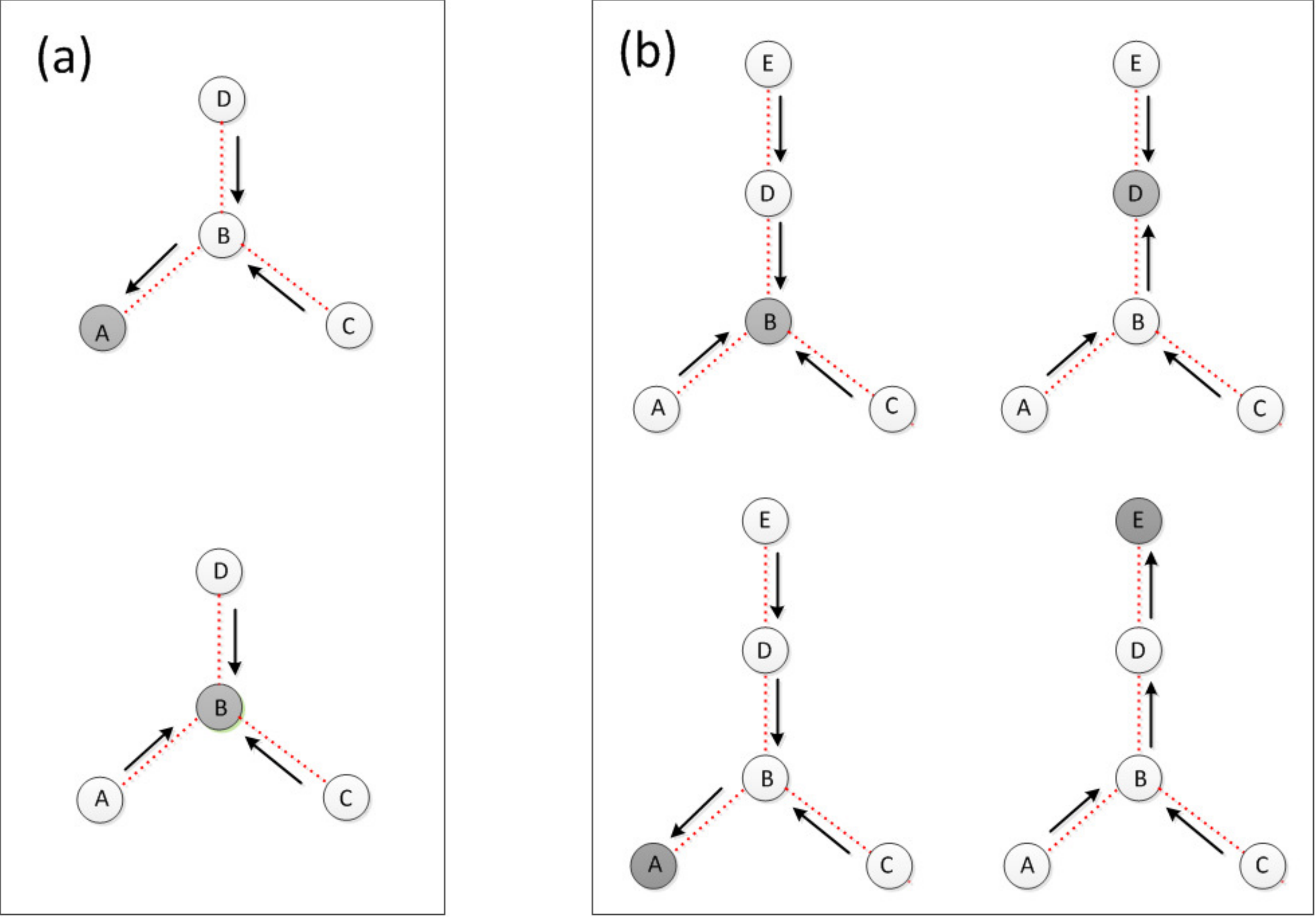}
\caption{ {\bf Schematic illustration of the level-by-level calculating process on a hierarchical structure} for (a) a star-like network of networks and (b) a tree-like network of networks. The gray circle denotes the network at the top level.}
\label{fig:LBL}
\end{figure}

For simplicity, we consider the case where the $M$ network layers within the multilayer system have an identical degree distribution: $p_{k}^{X}=p_{k}$.
We thus write $G^{X}_{0}(x)\equiv G_{0}(x)$, $G^{X}_{1}(x)\equiv G_{1}(x)$, and $\langle k\rangle^X\equiv z$ to simplify the notations. Assuming that the
network of $M$ network layers has no loop, we aim to obtain the equation governing the probability $S^{X}$ that a random node in a given layer $X$ belongs
to the giant component - the viable probability for this node. Suppose the node has $t$ failed replicas. Each link of this node is preserved with the probability $\alpha^{t}R^{X}$. The viable probability for this node is then $1-\sum_{k}p_k(1-\alpha^{t}R^{X})^{k}$. Let $f^{X}(t)$ be the probability distribution for the number of failed $t$ replicas of a random node in network layer $X$. We get the probability $S^{X}$ in terms of the generating function $G_0$ as
\begin{equation} \label{eq:S0}
S^{X}=\sum_{t=0}^{q^{X}}[1-G_{0}(1-\alpha^{t}R^{X})]f^{X}(t).
\end{equation}
From the probability distribution $f^{X}(t)$, we can also derive the equation for $R^{X}$ based on the branching process in network $X$. Following a randomly chosen link in layer $X$, we arrive at a node $(x,X)$ of degree $k$, where $k$ follows the distribution $kp_k/z$. With $t$ failed replicas, each link of the node $(x,X)$ is preserved with the probability $\alpha^{t}R^{X}$. The probability that node $(x,X)$ has at least one outgoing link to the giant component is
$1-(1-\alpha^{t}R^{X})^{k-1}$, which is also the probability that a randomly chosen link can lead to the giant component. Using the probability distribution $f^{X}(t)$ for a random node in layer $X$, we have
\begin{equation} \label{eq:R0}
R^{X}=\sum_{t=0}^{q^{X}}\alpha^{t}[1-G_{1}(1-\alpha^{t}R^{X})]f^{X}(t).
\end{equation}
To solve equations (\ref{eq:S0}) and (\ref{eq:R0}), we need the probability distribution function $f^{X}(t)$. We obtain a solution based on the level-by-level calculating process with respect to a hierarchical structure~\cite{Gleeson:2008,Gleeson:2007,Liu:2012}, as illustrated in Fig.~\ref{fig:LBL}. In particular, network layer $X$ is assigned to the top level, whose nearest neighbors constitute the $2$nd level, and the neighbors of the nearest neighbors belong to the $3$rd level, and so on. This way, for a given layer $Z$ with a super degree $q^{Z}$, it has $q^{Z}$ neighbors in the lower level if it is at the top level. Otherwise it has $q^{Z}-1$ neighbors in the lower level.

We first calculate the viable probabilities for the nodes in the bottom-level layers. For a random node in such a network layer $Z$, it has no lower-level
replicas. The probability distribution function $f^{Z}(t)$ for the number of failed replicas of a random node is given by
\begin{equation}
\label{cases}
f^{Z}(t)=\cases{1& if t=0,\\
0&otherwise.\\} \label{eq:f0}
\end{equation}
The viable probability $S^{Z}$ for a random node in network $Z$ is
\begin{equation} \label{eq:S}
S^{Z}=\sum_{t=0}^{q^{Z}-1}[1-G_{0}(1-\alpha^{t}R^{Z})]f^{Z}(t),
\end{equation}
where $q^{Z}-1$ is the number of neighboring networks in the lower level of network $Z$.

After calculating the viable probabilities for the nodes in all bottom-level network layers, we can get the probability distribution $f^{Y}(t)$ for a random node in a higher level network layer $Y$. Substituting $f^{Y}(t)$ into Eq.~(\ref{eq:S}), we can get the the viable probability $S^{Y}$. Repeating this process level by level, we can get the probability distribution function $f^{X}(t)$ for the top level.

\subsection{Star-like multilayer networks}

To be concrete, we consider a star-like interdependent system of four network layers, as shown in Fig.~\ref{fig:LBL}(a), where layer $B$ is the hub and the other three layers $A, C, D$ are peripheral networks (on the same footing). We thus have $R^{A}=R^{C}=R^{D}=R$ but $R^{B}=R'\ne R$. We first calculate the viable probability $S^{B}$ for a randomly chosen node in the hub layer $B$. The viable probability for a node in one of the peripheral (bottom-level) network layers is $1-G_0(1-R)$, and the probability distribution $f^{B}(t)$ for a random node in $B$ to have $t$ failed replicas is binomial: $f^{B}(t)=C_{3}^{t}G_{0}^{t}(1-R)[1-G_{0}(1-R)]^{3-t}$. From Eq.~(\ref{eq:S0}), we obtain the probability for a random node in layer $B$ to belong to its giant component:
\begin{equation} \label{eq:S1}
S^{B}=\sum_{t=0}^{3}C_{3}^{t}[1-G_{0}(1-\alpha^{t}R')]G_{0}^{t}(1-R)[1-G_{0}(1-R)]^{3-t}.
\end{equation}
Next, we calculate the viable probability for a random node in network $A$, which depends the state of its replica in layer $B$. If the replica is viable, we have $t = 0$; Otherwise $t = 1$. According to the updating process in Fig.~\ref{fig:LBL}, the viable probability of a node in layer $B$ is determined by the number $t$ of its failed replicas in the other two peripheral network layers (excluding layer $A$), which is
\begin{displaymath}
f^{A}(0)=\sum_{t=0}^{2}C_{2}^{t}[1-G_{0}(1-\alpha^{t}R')]G_{0}^{t}(1-R)
[1-G_{0}(1-R)]^{2-t}.
\end{displaymath}
The inviable probability for a random node in $B$ is thus given by $f^{A}(1)=1-f^{A}(0)$. From Eq.~(\ref{eq:S0}), we obtain the viable probability for a random node in layer $A$ as
\begin{equation} \label{eq:S2}
S^{A}=[1-G_{0}(1-R)]f^{A}(0)+[1-G_{0}(1-\alpha R)]f^{A}(1).
\end{equation}
The same equation holds for $S^{C}$ and $S^{D}$.

To solve Eqs.~(\ref{eq:S1}) and (\ref{eq:S2}) requires the equations for the quantities $R$ and $R'$, which we get through the branching process in network layers $A$ and $B$, respectively. In particular, for the probability distribution function $f^{B}(t)$, we arrive at a self consistency equation in terms of the generating functions:
\begin{equation} \label{eq:R1}
R'=\sum_{t=0}^{3}C_{3}^{t}\alpha^{t}[1-G_{1}(1-\alpha^{t}R')]
G_{0}^{t}(1-R)[1-G_{0}(1-R)]^{3-t}.
\end{equation}
Similarly, the self consistency equation for $R$ is
\begin{equation} \label{eq:R2}
R=[1-G_{1}(1-R)]f^{A}(0)+\alpha[1-G_{1}(1-\alpha R)]f^{A}(1).
\end{equation}
It is not feasible to get closed-form expressions for $R'$ and $R$ from Eqs.~(\ref{eq:R1}) and (\ref{eq:R2}). We thus resort to numerical solutions. For a given degree distribution $p_{k}$ and a fixed value of the tolerance parameter $\alpha$, we plot the curves for Eqs.~(\ref{eq:R1}) and (\ref{eq:R2}) on the $(R-R')$ plane, where the coordinates for the top crossing point give the solutions. Substituting the values of $R'$ and $R$ into Eqs.~(\ref{eq:S1}) and (\ref{eq:S2}), we can get the sizes of the giant components $S^{A}$ and $S^{B}$.

\begin{figure}
\centering
\includegraphics[width=\linewidth]{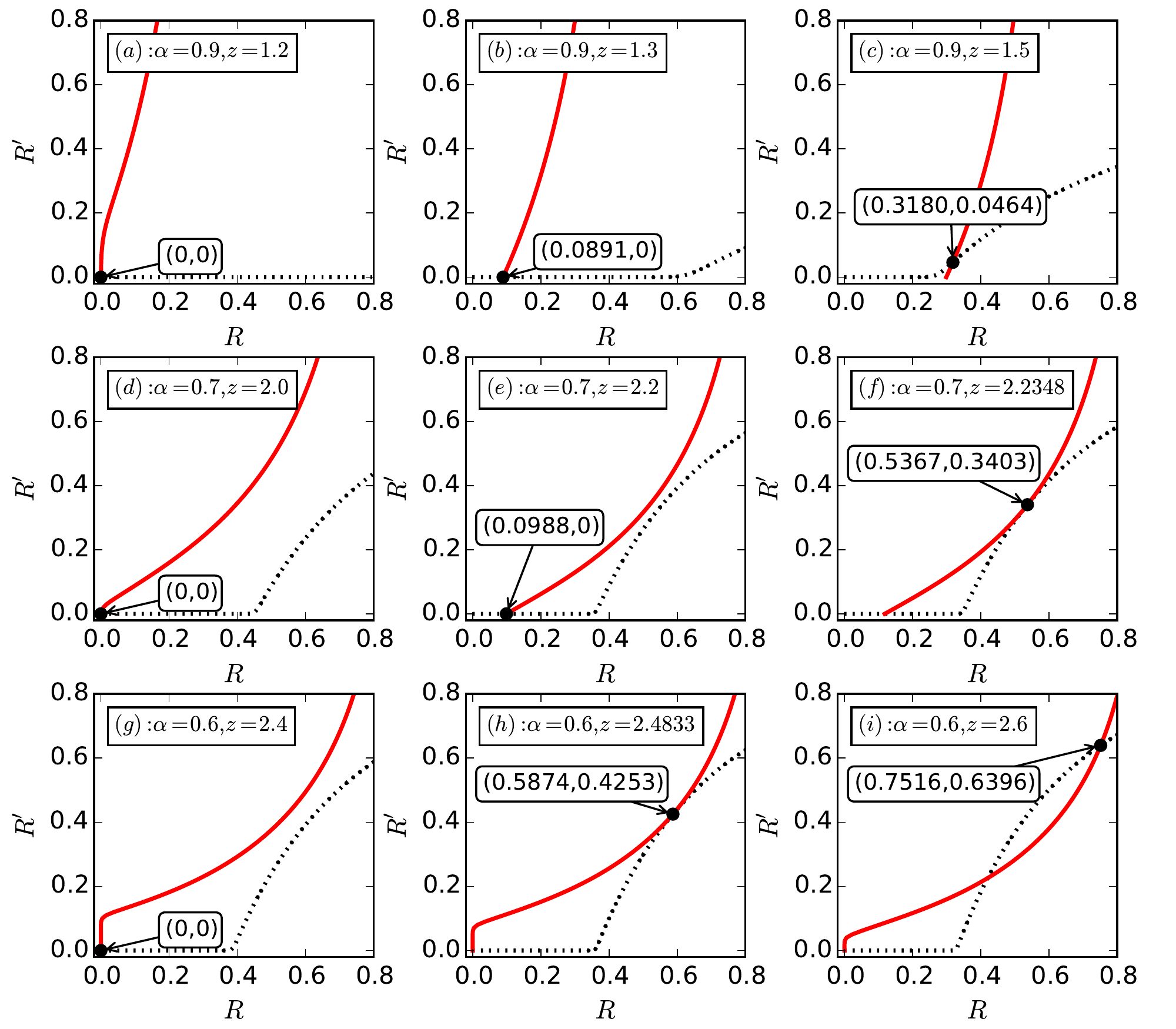}
\caption{ {\bf Percolation transitions in the multilayer star system of four network layers}.
Shown are graphical solutions of Eqs.~(\ref{eq:R1}) and (\ref{eq:R2}) for
different values of $\alpha$ and $z$, as marked by the black dots.
(a-c) For $\alpha = 0.9$, the results for $z=1.2$, $z=1.3$ and $z=1.5$,
respectively. (d-f) For $\alpha = 0.7$, results for $z=2$, $z=2.2$, and
$z=2.2348$, respectively. (g-i) For $\alpha = 0.6$, the solutions for
$z=2.4$, $z=2.4833$, and $z=2.6$, respectively.}
\label{fig:Star}
\end{figure}

\begin{figure}
\centering
\includegraphics[width=\linewidth]{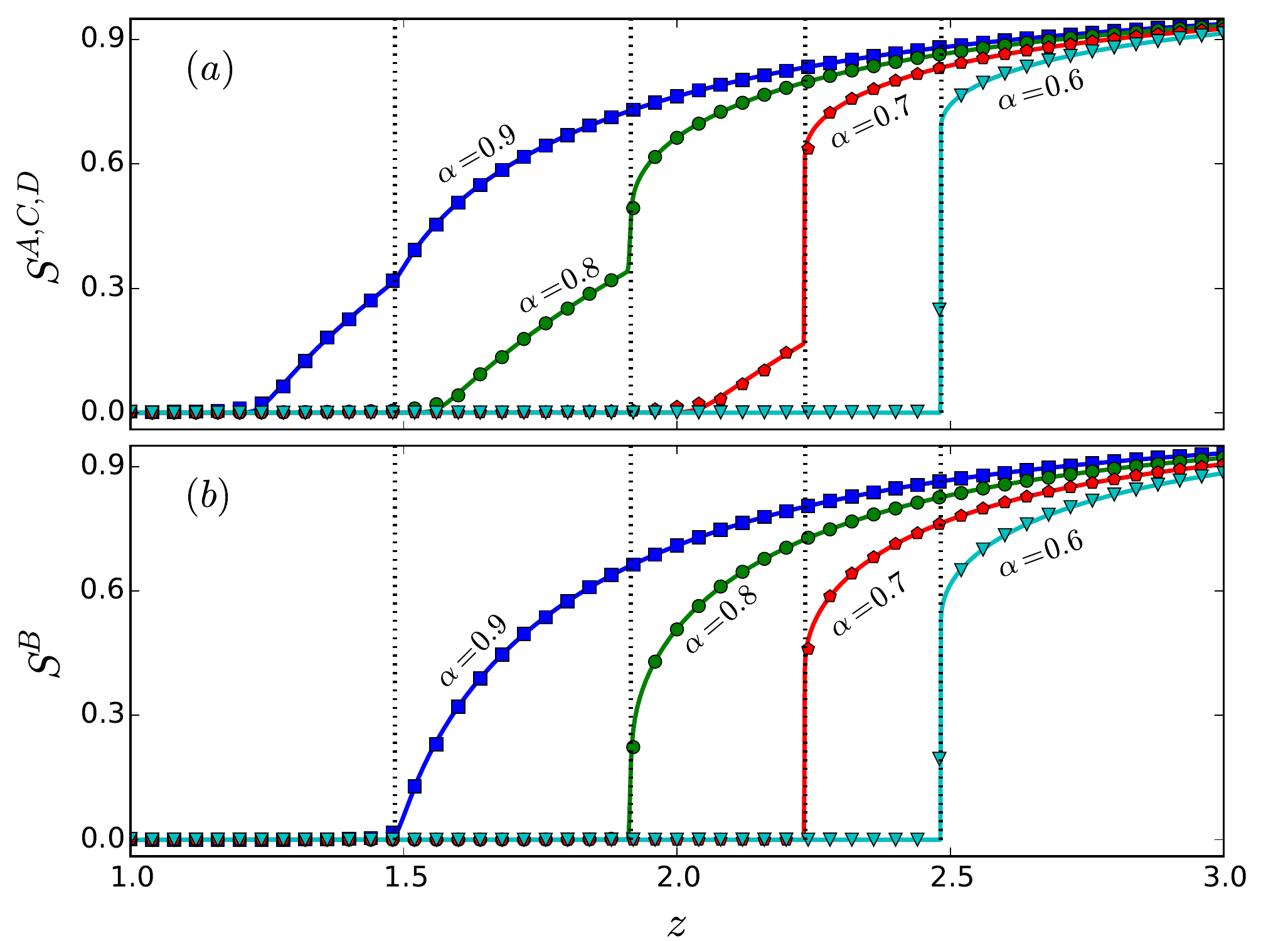}
\caption{ {\bf Rich percolation transition behaviors in the interdependent
star system of four random network layers with an identical degree distribution}.
For different values of the tolerance parameter $\alpha$, (a) the sizes of
the giant components, $S^A$, $S^C$, $S^D$ versus the Poisson degree
distribution parameter $z$, and (b) the size of the giant component
$S^B$ versus $z$, where the dotted vertical lines mark the percolation
threshold for $B$. The solid curves are theoretical predictions while the
symbols are direct simulation results. The network size is $N = 10^5$. Each
data point is the result of averaging over 20 independent statistical
realizations. The remarkable phenomenon of mixed percolation transitions
occurs for intermediate $\alpha$ values (e.g., $\alpha = 0.8$ and
$\alpha = 0.7$), where the hub network layer exhibits a discontinuous transition
but each peripheral network layer exhibits double transitions: one continuous and
another discontinuous.}
\label{fig:Star_GC}
\end{figure}

Figure~\ref{fig:Star} shows, for random networks with a Poisson distribution~\cite{Bollobas:1985} $p_k=e^{-z}z^k/k!$, graphical solutions of $R$ and $R'$ for different values of $\alpha$ and $z$. For $\alpha=0.9$, there is a trivial solution at the point $(R=0,R'=0)$ for $z=1.2$, indicating that the hub and the peripheral network layers are completely fragmented. For $z=1.3$, a nontrivial value of $R$ arises but $R'$ remains to be zero, suggesting a continuous phase transition for $R$. Increasing $z$ to $1.5$, we observe that $R'$ also becomes nontrivial which changes growth rate for $R$. These results imply that, for $\alpha = 0.9$, the peripheral network layers percolate firstly, and then the hub percolates as $z$ is increased, leading to double phase transitions for the peripheral network layers. For $\alpha=0.7$, the phenomenon of mixed percolation transitions occur, where the peripheral network layers percolate in a continuous manner, as shown in Fig.~\ref{fig:Star}(e), but the hub percolates in a discontinuous fashion, as demonstrated by the tangent point shown in Fig.~\ref{fig:Star}(f). Specifically, for $z < 2.2348$, the solutions for $(R,R')$ are given by the crossing point of the curves with the $R$-axis. However, for $z \approx 2.2348$, their solutions are given by the tangent point $(0.5367,0.3403)$, giving rise to a discontinuous change in both $R$ and $R'$. For $\alpha=0.6$, we find that the crossing point for solutions of $R$ and $R'$ change abruptly from $(0,0)$ to the tangent point $(0.5874,0.4253)$, indicating also a discontinuous percolation
transition.

The sizes $S^A$, $S^B$, $S^C$, and $S^D$ of the giant components for the four network layers in the star system versus the average degree $z$ are shown
in Fig.~\ref{fig:Star_GC}. (See Sec.~\ref{sec:simulation} for an explanation of the numerical methods.) For $\alpha = 0.9$, as $z$ is increased from the
value of one, all four layers exhibit a continuous percolation transition, where the peripheral layers (layers A, C, and D) percolate first at the critical point $z_{c1} \approx 1.2345$ [Fig.~\ref{fig:Star_GC}(a)], followed by the hub network (layer B) at the critical point $z_{c2} \approx 1.4872$, as shown in Fig.~\ref{fig:Star_GC}(b). We see that, while the curves of $S^A$, $S^C$, and $S^D$ versus $z$ are continuous, they exhibit a discontinuity in their derivative at both $z_{c1}$ and $z_{c2}$. In this sense we say that the peripheral network layers exhibit double percolation transitions. Note that, in the whole range of $z$ values considered, the hub layer exhibits only a single percolation transition - continuous (or second order) for relatively large values of $\alpha$ but discontinuous (of first order) for smaller $\alpha$ values. The feature of double percolation transitions for the peripheral network layers persists for
$\alpha = 0.8$ and $\alpha = 0.7$. Remarkably, for these two values of $\alpha$, the two percolation transitions for the peripheral networks are characteristically different: the first one is continuous while the second is discontinuous. In contrast, the transition for the hub network layer is discontinuous. Thus, from the standpoint of the whole interdependent, multilayer system, for these values of $\alpha$, both continuous and discontinuous percolation transitions exist, leading to the phenomenon of mixed percolation transitions. For $\alpha=0.6$, the hub and peripheral layers percolate discontinuously at the same point. These results indicate a richer variety of percolation transition scenarios than revealed in previous studies for multilayer systems with strong interdependence (i.e., $\alpha = 1$ ~\cite{Gao:2011}).

A practical implication is that the tolerance parameter can be exploited for modulating or controlling the characteristics of the percolation transition. In particular, for relatively small values of $\alpha$ (e.g., $\alpha = 0.6$), the percolation transitions are abrupt and discontinuous - the defining characteristic of first-order phase transitions. In this case, the interdependent system will collapse suddenly as a system parameter is changed in response to random nodal failures or intentional attacks. The system is thus not resilient. To improve the resilience of the system, a larger value of $\alpha$ can be chosen (e.g., $\alpha = 0.9$). In this case, the percolation transitions in the interdependent networks are continuous - signature of second-order phase transitions. While the whole system still collapses, the manner by which the collapse occurs is benign as compared with that of first-order phase transitions.

\subsection{Tree-like multilayer networks}

\begin{figure}
\centering
\includegraphics[width=\linewidth]{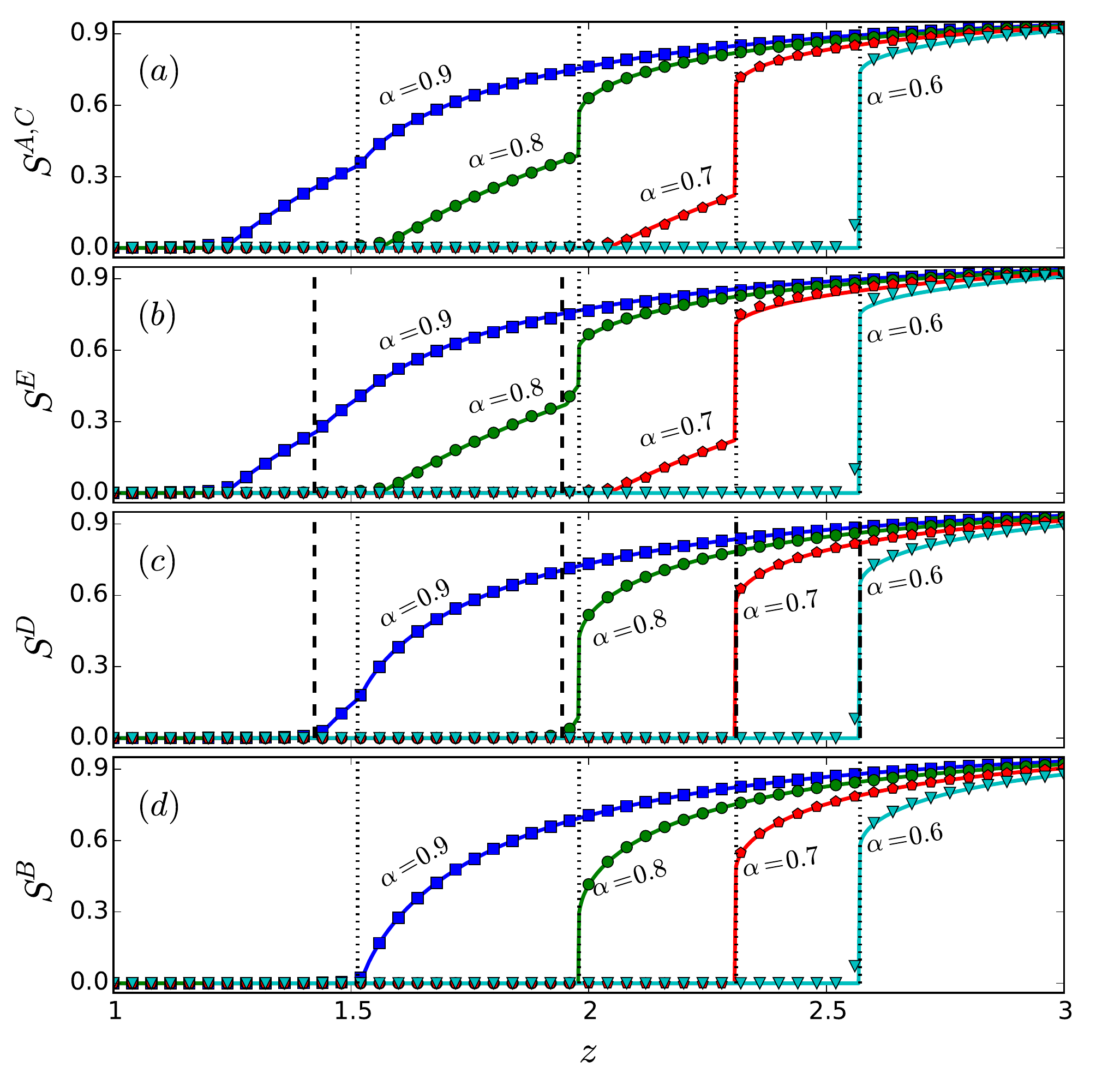}
\caption{ {\bf Second-order and first-order percolation transitions for a tree-like interdependent system of five random network layers}. Shown are the sizes of the giant components $S^A$ or $S^C$ (a), $S^E$ (b), $S^D$ (c) and $S^B$ (d) versus the random-network parameter $z$, where the solid lines represent theoretical predictions and the symbols are numerical results averaged over $20$ independent network realizations. The network size is $N = 10^5$. The dotted vertical lines mark the percolation thresholds for $B$ or the phase transition points for other layers induced by its percolation, and the dashed vertical lines mark the percolation thresholds for $D$ or the induced phase transition points for other networks.}
\label{fig:Tree_GC}
\end{figure}

We consider a tree-like interdependent system of five random network layers, labeled as $A,B,C,D$ and $E$ respectively, as shown in Fig.~\ref{fig:LBL}(b), and obtain theoretical solutions for the sizes of the giant components and for the nodal viable probabilities for all the layers. Figure~\ref{fig:Tree_GC} shows the theoretical and numerical solutions of the sizes of giant components $S^{A}, S^{B}, S^{C}, S^{D}$ and $S^{E}$ versus the average degree $z$ (see Sec.~\ref{sec:simulation} for an explanation of the numerical methods). For $\alpha=0.9$, we find that the peripheral layers $A,C,E$ percolate first, followed by the sub-center layer $D$. The size of the giant component of $D$'s nearest neighboring layer $E$ is then increased, leading to a continuous phase transition for $S^{E}$. The central network layer $B$ percolates last, giving rise to a distinct continuous transition for its nearest neighboring networks $A,C,D$. As $\alpha$ is decreased, the hub network percolates discontinuously, at which the giant components of layers $A, C, E$ and $D$ increase abruptly in size, giving rise to the phenomenon of mixed percolation transitions, as can be seen from the curves for $\alpha=0.8$. For $\alpha = 0.7$, the percolation transition points for some network layers with large super degrees begin to merge but the phenomenon of mixed percolation transitions persists. As $\alpha$ is decreased further, a first-order phase transition occurs at which all layers percolate at the same point in a discontinuous manner, as indicated by the curves for $\alpha=0.6$.

These transition phenomena suggest that the super degree of a layer in the multilayer system plays a critical role in its percolation transition. In particular, the peripheral network layers with the lowest super degree percolate first, followed by the layers with a moderate value of the super degree, and finally by the layers with the highest super degree. When a layer with a larger super degree percolates continuously, it will increase the giant component sizes of its neighboring network layers that have already percolated, leading to multiple percolation transitions. In contrast, if a layer with a larger super degree percolates discontinuously, it will lead to a sudden and discontinuous increase in the sizes of the giant components of all network layers that have already percolated, leading to mixed percolation transitions. These results indicates that the percolation type of the hub layers plays a critical role in the robustness of the whole system. In particular, if the hub layer percolates continuously, at the transition point the sizes of the giant components of the nearest network layers are continuous but their derivatives are discontinuous. However, if the hub layer percolates discontinuously, an abrupt and discontinuous increase in the giant component sizes of the neighboring network layers can occur.

\section{Direct simulation results} \label{sec:simulation}

To verify the phenomena of mixed and multiple percolation transitions directly, we perform bond percolation $20$ times using the Newman-Ziff algorithm~\cite{Newman:2000} and measure the average relative size of the viable component of each network layer. As shown in Figs.~\ref{fig:Star_GC} and \ref{fig:Tree_GC}, the percolation transitions and the behaviors of the giant components as predicted theoretically agree well with the numerical results.

\begin{figure}
\centering
\includegraphics[width=\linewidth]{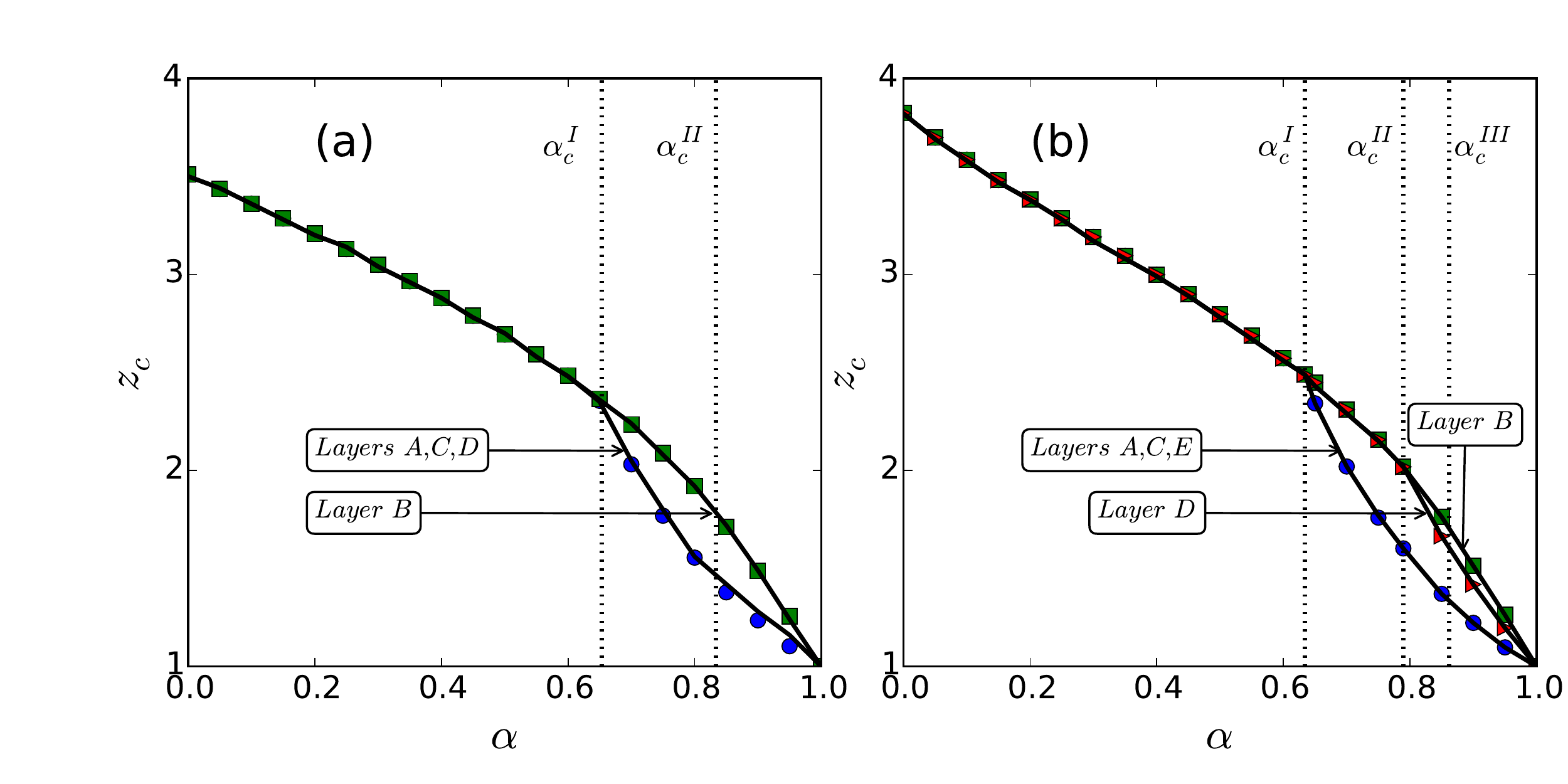}
\caption{ {\bf Dependence of the percolation transition point on the tolerance parameter}. (a) For the star-like system, there are two critical points: $\alpha_{c}^{I}\approx 0.6513$ and $\alpha_{c}^{II}\approx 0.8337$, which divides the $\alpha$ interval into three subregions with distinct transition behaviors. (b) For the tree-like system, there are three critical points: $\alpha_{c}^{I}\approx 0.6339$, $\alpha_{c}^{II}\approx 0.7894$ and $\alpha_{c}^{III}\approx 0.8614$, which divide the $\alpha$ interval into four subregions. The solid lines represent the theoretical predictions and the symbols are simulation results based on the Newman-Ziff bond percolation algorithm. Each data point is the averaging result of 20 network realizations. The network size is $N = 10^5$.}
\label{fig:transition}
\end{figure}

Figures~\ref{fig:transition}(a) and \ref{fig:transition}(b) show, for the star-like and tree-like interdependent systems, respectively, the percolation transition points $z_{c}$ versus $\alpha$ for each layer in the multilayer system, which are obtained by the graphical solution as illustrated in Fig.~\ref{fig:Star}. An alternative way to identify the transition points is to examine the fluctuations in the size of the giant component which, for a finite system, tend to be relatively large at the transition~\cite{Zhou:2014}. From Fig.~\ref{fig:transition}(a), we see that the phase diagram for the star-like system can be divided into three different regions by the two critical points denoted as $\alpha_{c}^{I}$ and $\alpha_{c}^{II}$. For $\alpha\in[\alpha_{c}^{II}, 1]$, both the hub layer $B$ and the peripheral layers $A, C, D$ percolate in a continuous manner (second-order phase transition) but at different transition points. For $\alpha_{c}^{I} \le \alpha < \alpha_{c}^{II}$, the peripheral layers first percolate in a continuous fashion, and then the hub layer percolates discontinuously (first-order phase transition), leading to mixed percolation transitions. For $\alpha\in[0,\alpha_{c}^{I})$, all network layers percolate discontinuously at the same transition point. Similarly, for the tree-like system, Fig.~\ref{fig:transition}(b) indicates a division of the $\alpha$ interval into four subregions determined by the critical points $\alpha_{c}^{I}$, $\alpha_{c}^{II}$ and $\alpha_{c}^{III}$. For $\alpha\in(\alpha_{c}^{III},1]$, all layers percolate continuously: the peripheral network layers $A,C,E$ percolate first at the same transition point, followed by the percolation of the sub-central layer $D$ and finally by the central layer $B$. For $\alpha\in(\alpha_{c}^{II},\alpha_{c}^{III}]$, the phenomenon of mixed percolation transitions arises. That is, the percolation transition for the central layer $B$ is discontinuous and those for network layers $A,C,D,E$ remain to be continuous. For $\alpha_{c}^{I}< \alpha\le \alpha_{c}^{II}$, the phenomenon of mixed percolation transitions persists. The sub-central layer $D$ and the central layer $B$ percolate simultaneously at a common transition point, and the peripheral layers $A,C,E$ still percolate at the same point in a continuous manner. For $0 \le \alpha \le \alpha_{c}^{I}$, all network layers percolate discontinuously at the same transition point.

\section{Discussion} \label{sec:discussion}

The results of this paper demonstrate that the pervasive assumption of ``strong" interdependence in percolation models is limiting our understanding of the robustness of multilayer systems. When tuning the interdependence between networks via a cascade tolerance mechanism, the surprising phenomenon of mixed percolation transitions occurs: network layers with large super degrees percolate discontinuously while those with small super degrees percolate continuously. That is, multilayer networks with weak interdependence across layers, represented by moderate values of the tolerance parameter, experience both second-order and first-order percolation transitions. This result is unlike those in any study that assumes either strong or no interdependence between network layers, where all network layers experience either simultaneous and abrupt percolation or continuous percolation, respectively. These previous models imply that catastrophic cascades were either inevitable or impossible, yet our results demonstrate that both the strength of interdependence and layer position play a critical role in the functioning of interdependent systems. In cases where the connectivity of the central layer is critical to the functioning of the interdependent system, sufficient tolerance must be included to prevent sudden, system-wide failures. This result implies that studies that do not consider interdependence may underestimate the amount tolerance needed to ensure cascades do not occur. On the other hand, where layers with high super degree only serve as bridges to critical periphery layers, far less tolerance is needed to maintain interdependent function. Likewise, studies that assume strong interdependence may be underestimating the likelihood that the system will survive. In both cases, the under- and over-estimation of multilayer network robustness may produce unwanted cascading dynamics if used to study and design real networks.

Since cascade tolerance provides a generic metric for the physical, temporal, and dynamic buffers found in real infrastructure to prevent losses across interdependent systems, the results of this paper have broad implications for infrastructure design. Infrastructure design often uses risk analysis to give the individual assets that comprise large-scale infrastructure systems tolerances to known failure modes. For example, individual assets have an oversized capacity to handle sudden shifts in service flows, have uninterruptible power supplies to ensure asset functioning during blackouts, and can be built a certain number of feet above the ground to prevent flooding. Despite significant work protecting individual assets in infrastructure systems, there is far less understanding how risk adverse practices in a single facility may impact the robustness of an entire infrastructure system or interdependent, multilayered system. The tolerance parameter introduced in our model is generic to capture the probability that these risk adverse practices influence percolation dynamics in a wide variety of infrastructure systems. Although previous models would indicate that interdependent systems could experience more sudden, global cascades, mixed percolation transitions may be an indicator of an even more precarious situation. With buffers that appear sufficient within a single network layer, a central layer with high super degree can still experience sudden, catastrophic failures that inhibits functionality of the multilayer network. This situation is common in disaster situations, where individual assets and entire infrastructure systems may survive the initial failure event, but remain unusable because interdependent systems that they require to function did not. The values of $\alpha$ that generate mixed percolation transitions in our model provide a heuristic measure of when these interdependent failures may occur that is characteristically different from other percolation models. Thus, our results suggest that developing a metric for buffering capacity and identifying the super degree relationships among infrastructure systems may reveal a disconnect between local, risk adverse practices, and global, interdependent vulnerability.

Although mixed percolation transitions have important implications for future percolation models and infrastructure design, we do concede that our current model is too simplified to represent many real-world systems. Percolation dynamics provide an important, generic method for considering the interactions within multilayer networks. However, the dynamics that occur within real-world systems are not captured in this model, suggesting that our assumptions for tolerance and cascading dynamics remain unrealistic. For example, modeling the failure dynamics in a multi-modal transportation network like the one discussed above (see Section 2) requires detailed information regarding the infrastructure located within each city and models for how people choose between different transportation modes. The current percolation model assumes that the loss of an airport in a periphery layer affects losses in interdependent layers equally, when real travel information may show greater heterogeneity in congestion and failures. Future work should focus on linking discipline-specific infrastructure models that capture these dynamics with measures of cascade tolerance across network layers to produce more realistic percolation transitions.

\ack
RRL was supported by the National Natural Science Foundation of China under Grant Nos.~61773148, 61403114 and 61673150, and by the Zhejiang Provincial
Natural Science Foundation of China under Grant No.~LQ14F030009. DAE, TPS, and YCL were supported by NSF under Grant No.~1441352. DAE was also supported by NSF Grant No.~1311230 the Defense Threat Reduction Agency Grant No.~DTRA-18681. YCL would also like to acknowledge support from the Vannevar Bush Faculty Fellowship program sponsored by the Basic Research Office of the Assistant Secretary of Defense for Research and Engineering and funded by the Office of Naval Research
through Grant No.~N00014-16-1-2828.

\section*{References}

\newcommand{\noop}[1]{}
\providecommand{\newblock}{}

\end{document}